\documentclass[prl,letterpaper,twocolumn,superscriptaddress,floatfix]{revtex4-2}

\usepackage{graphicx,psfrag,amsmath,amssymb,amsfonts,bbm,latexsym,color,dcolumn,bm,mathbbol}
\usepackage[framemethod=tikz]{mdframed}
\usepackage{hyperref}
\hypersetup{
colorlinks=true,
linkcolor=blue,
citecolor=blue,
filecolor=blue,      
urlcolor=blue,
pdftitle={Overleaf Example},
pdfpagemode=FullScreen,
}

\usepackage[utf8]{inputenc}
\usepackage{natbib}
\usepackage{graphicx}

\usepackage{xcolor}

\usepackage{comment}

\usepackage{soul} 
\usepackage{cancel}

\usepackage{array}
\usepackage{booktabs}
\usepackage{footnote}
\usepackage{threeparttable}

\usepackage{natbib}
\usepackage{amsmath,amssymb}             % AMS Math
\usepackage{color}

\usepackage{enumerate}
\usepackage[normalem]{ulem}
\definecolor{jfcolor}{rgb}{0.1, 0.0, 0.9}

\definecolor{szlcolor}{rgb}{0.75, 0, 0.75}

\begin{document}

% \graphicspath{{Figures/}}

\title{Spontaneous flows in active smectics with dislocations}

\date{\today}

\author{Shao-Zhen Lin}
\affiliation{Aix Marseille Univ, Université de Toulon, CNRS, CPT (UMR 7332), Turing Centre for Living Systems, Marseille, France}
\author{Frank J\"ulicher}
\affiliation{Max Planck Institute for the Physics of Complex Systems, 01187 Dresden, Germany}
\affiliation{Cluster of Excellence, Physics of Life, TU Dresden, 01307 Dresden, Germany}
\author{Jacques Prost}
% \email{jacques.prost@curie.fr}
\affiliation{Laboratoire Physico-Chimie Curie, UMR 168, Institut Curie, PSL Research University, CNRS, Sorbonne Université, 75005 Paris, France}
\affiliation{Mechanobiology Institute, National University of Singapore, 117411 Singapore}
\author{Jean-François Rupprecht}
% \email{jean-francois.rupprecht@univ-amu.fr}
\affiliation{Aix Marseille Univ, Université de Toulon, CNRS, CPT (UMR 7332), Turing Centre for Living Systems, Marseille, France}

\begin{abstract} 
We construct a hydrodynamic theory of active smectics A in two-dimensional space, including the creation/annihilation and motility of dislocations with Burgers’ number $\pm 1$. We derive analytical criteria on the set of parameters that lead to flows. We show that the motility of dislocations can lead to flow transitions with distinct features from the previously reported active Helfrich--Hurault shear instability with, notably, a discontinuous transition in the velocity from quiescence to turbulence.
\end{abstract}
%We discuss connections with the case without dislocation. Our work paves the way for a generic quantitative understanding of the large-scale properties of active layered systems.  

\maketitle

\textit{Introduction.} -- Smectics are systems characterized by the presence of a one-dimensional modulation and fluid order in the direction orthogonal to the modulation. Typically, they are made of stacks of molecular layers, each layer being fluid, like in hydrated soaps \cite{de1993physics}. 
Smectic hydrodynamics describe the long-wavelength, long-time dynamical behavior of such systems. It was extended to include local energy consumption \cite{Adhyapak_PRL_2013}, with a recent work exploring the mechanical conditions for the onset of active flows \cite{Kole_PRL_2021}. These generic theories apply to non-living, out-of-equilibrium systems (assemblies of organic molecules in a temperature gradient \cite{Podoliak_PRL_2023}, or Rayleigh-Bénard type convection rolls, with each roll playing the role of a layer \cite{PhysRevB.23.316}) and living materials, e.g., motor microtubule gels \cite{Senoussi2019} or actomyosin structures in lamellipodia \cite{Hu2017,Seul_Science_1995}. 

\begin{figure}[t!]
\includegraphics[width=8.7cm]{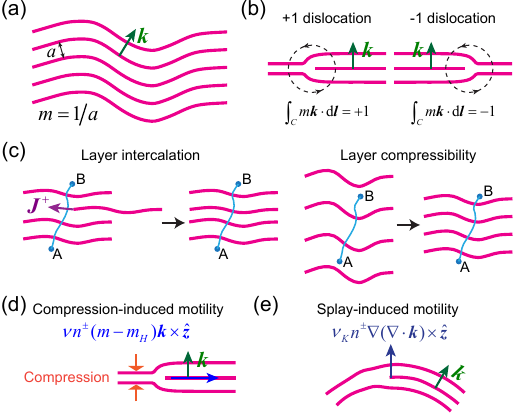}
\caption{Model sketch. 
(a) Smectic layers with the definition of $m$ and $\bm{k}$. 
(b) Sketch of a $+1$ dislocation and a $-1$ dislocation. 
(c) Sketch of the two contributions to the layer number variation between the positions A and B. (\textit{left}) Layer intercalation induced by dislocation flux; (\textit{right}) Layer compressibility. 
(d, e) Sketches of the compression-induced motility (light blue arrow) (d) and splay-induced motility (dark blue arrow) (e) of $+1$ dislocations.  
}
\label{fig:ModelSketch}
\end{figure}

\textit{Theory.} -- In this paper, we focus on smectics A, composed of stacks of orientationally aligned molecules, 
in a two-dimensional setting (Fig. \ref{fig:ModelSketch}(a)). 
We describe the active smectic through five fields: (1) $\bm{k}$, a unit vector field describing the normal to the layers; (2) $m$, a scalar describing the inverse interlayer distance ($m=1/a$ \cite{Julicher_PRE_2022}); %it corresponds to the local number of layers per unit length normal to the layers ; 
(3) $\bm{v}$, the flow velocity; (4) $n^{+}$ and (5) $n^{-}$ are the densities in $+1$ and $-1$ dislocations, respectively (Fig. \ref{fig:ModelSketch}(b)). Next, we derive the governing equations for such a system. 

\textit{Layer number balance.} We first derive the dynamic evolution equation of $\bm{M} = m\bm{k}$, called smectic director. Considering two points A and B, see Fig. \ref{fig:ModelSketch}(c), the number of layers crossed along a path between A and B reads $N_{A \rightarrow B} = \int_{A}^{B} m \bm{k} \cdot \mathrm{d}\bm{l}$ \cite{Julicher_PRE_2022}. 
Over a closed path, $A=B$ and $\oint m \bm{k} \cdot \mathrm{d}\bm{l} = \int_{S} (n^{+} - n^{-}) \mathrm{d}S$, hence, applying Stokes' theorem, we obtain that:
\begin{equation}
\nabla \times \left( m\bm{k} \right)=\left(  {{n}^{+}}-{{n}^{-}}\right) \hat{\bm{z}}, \label{eq:TopologicalRelation}
\end{equation}
with $\hat{\bm{z}}$ a unit vector normal to the plane. 
The temporal variation, $\mathrm{d}N_{A \rightarrow B} / \mathrm{d}t = \int_{A}^{B} {\partial_t (m\bm{k})} \cdot \mathrm{d}\bm{l}$, includes two contributions: (1) the layer intercalation due to dislocation flux, $\int_{A}^{B} [-\hat{\bm{z}} \times (\bm{J}^{+} - \bm{J}^{-})] \cdot \mathrm{d}\bm{l}$ with $\bm{J}^{\pm}$ the flux of $\pm 1$ dislocations, see Fig. \ref{fig:ModelSketch}(c); (2) the layer compressibility, expressed as $J_m^{(A)} - J_m^{(B)} = - \int_{A}^{B} \nabla J_m \cdot \mathrm{d}\bm{l}$, with $J_m$ a potential function that effectively describes flux of smectic layers in the normal direction, see Fig. \ref{fig:ModelSketch}(c). Therefore, $\int_A^{B} {\partial_t (m\bm{k})} \cdot \mathrm{d}\bm{l} = -\int_A^B [ \hat{\bm{z}} \times (\bm{J}^{+} - \bm{J}^{-}) + \nabla J_m ] \cdot \mathrm{d}\bm{l}$, for all paths $\mathrm{A}\rightarrow\mathrm{B}$; this leads to following layer number balance equation: 
\begin{equation}
{\partial_t (m\bm{k})} + \nabla {{J}_{m}} = - \hat{\bm{z}}\times ( {{\bm{J}}^{+}}-{{\bm{J}}^{-}} ) . \label{eq:mk_evolution}
\end{equation}
%and $\pm 1$ dislocations $\bm{J}^{\pm}$ 
The flux of smectic layers $J_m$ can be expressed as
\begin{align}
\hskip-0.5cm J_{m} = m\bm{v}\cdot \bm{k}-{{D}_{m}}\bm{k}\cdot \nabla m+{{\lambda }_{m}}\nabla \cdot \bm{k} + \xi_m \nabla^2(\nabla\cdot\bm{k}), 
\label{eq:Jm}
\end{align}
which obeys up-down and rotational symmetry and describes the layer advection, layer elasticity, and activity ($\lambda_m$, called $D$ in Refs. \cite{Adhyapak_PRL_2013, Ramaswamy_PRL_2000}), as well as the layer bending rigidity \cite{Julicher_PRE_2022}. In turn, the fluxes $\bm{J}^{\pm}$ of $\pm 1$ dislocations read
\begin{align}
{{\bm{J}}^{\pm}} = \ & {{n}^{\pm}}\bm{v}-D_n\nabla n^{\pm}\pm\nu{{n}^{\pm}}\left( m-{{m}_{H}} \right) \bm{k}\times \hat{\bm{z}} \notag \\ 
& \pm{{\nu }_{K}}n^{\pm}\left( \nabla \left( \nabla \cdot \bm{k} \right)\times \hat{\bm{z}} \right), \label{eq_J_positive}
\end{align} %  the $\nu$ term refers to 
where each term represents, respectively, the advection and diffusion of dislocations, the Peach-Koehler compression-induced motility \cite{Peach1950} ($\nu \geq 0$; see Supplemental Material \cite{SM}, Sec. I), and a splay-induced motility of dislocations \cite{Pershan1975} (with a preference to high splay zones when $\nu_K>0$).

\textit{Dislocation dynamics.} Annihilation and creation of dislocations occur by pairs. We thus consider the following dynamic evolution of the number density of dislocations: 
\begin{equation}
{\partial_t {{n}^{\pm}}} +\nabla \cdot {{\bm{J}}^{\pm}} = -\alpha{{n}^{+}}{{n}^{-}} + \beta. \label{eq:n_pm_evolution}
\end{equation}
where $\beta > 0$ is a nucleation rate per area and $\alpha > 0$ describes annihilation. This set of equations is consistent with Eq. (\ref{eq:TopologicalRelation}); see Supplemental Material \cite{SM}, Sec. I. 

\textit{Force balance equation.} The force balance equation
\begin{equation}
\nabla \cdot \bm{\sigma }-\lambda\bm{v} -{{\zeta }_{s}} ( \nabla \cdot \bm{k} ) \bm{k}-{{\zeta }_{b}}\bm{k}\cdot \nabla \bm{k} = \bm{0}, \label{eq:ForceBalance}
\end{equation}
relates the divergence of the stress tensor ($\bm{\sigma}$) within the active smectic layers to the momentum exchange with the underlying substrate. This includes passive friction ($\lambda$) and active traction induced by splay ($\zeta_s$) and bend ($\zeta_b$) deformations \cite{Duclos2018,Maitra2018}. 
We consider $\bm{\sigma } = - P\bm{I} + 2\eta\bm{E} + \bm{\sigma }_B + \bm{\sigma }_K$, where $\eta$ is the viscosity, $\bm{E} = [ \nabla\bm{v} +(\nabla\bm{v})^T ] / 2$ the strain-rate, and $P$ is the pressure acting as Lagrange multiplier imposing the incompressibility constraint 
\begin{equation}
    \nabla \cdot \bm{v} = 0,
\end{equation} 
%The parameter $m_H$ refers to a reference interlayer spacing density at equilibrium, while $m_S$ corresponds to a homeostatic value. 
while $\bm{\sigma}_B = - B [({m-{{m}_{S}}})/{{{m}_{H}}}]( \bm{k}\otimes \bm{k}-\bm{I}/2)$ with $B$ the layer compressibility, $m_H$ the homeostatic level of layer spacing density \cite{Julicher_PRE_2022}, and $m_S $ the stress-free layer density for which the normal stress component vanishes. For a passive system, $m_S = m_H$, yet this is not imposed in the active case. Finally, $\bm{\sigma}_K = K [ \bm{k}\otimes \nabla ( \nabla \cdot \bm{k} )+\nabla ( \nabla \cdot \bm{k} )\otimes \bm{k}-\bm{k}\cdot \nabla ( \nabla \cdot \bm{k} )\bm{I} ]$ is the stress associated with large bending, with $K$ the layer bending rigidity \cite{Adhyapak_PRL_2013,Kole_PRL_2021}. 

\textit{Linear stability analysis.} -- We investigate the dynamic behavior of the set of Eqs. \eqref{eq:mk_evolution}, \eqref{eq:n_pm_evolution}, and \eqref{eq:ForceBalance} starting with the stability for all wavevectors (i.e., infinitely large system) of the homogeneous state: $m(\bm{x}) = m_H$, $\bm{k}(\bm{x}) = \hat{\bm{y}}$, ${{n}^{+}}(\bm{x})={{n}^{-}}(\bm{x}) = n_H$ with $n_H = 2 \sqrt{\beta / \alpha}$, and $\bm{v}(\bm{x}) = \bm{0}$. The perturbation equation 
reads: $\partial_t \widetilde{\delta \bm{M}} = \bm{G} (\bm{q})\cdot \widetilde{\delta \bm{M}}$, 
where $\bm{q}$ is the wavevector, $\widetilde{\delta \bm{M}}$ is the Fourier transform of the perturbation $\delta \bm{M} = \bm{M} - m_H \hat{\bm{y}}$, and $\bm{G} (\bm{q})$ is the Jacobian matrix. %; see Supplemental Material \cite{SM}, Sec. I. 
The stability of the homogeneous state is determined by the eigenvalues $\omega_{\pm} (\bm{q}) = (-b \pm \sqrt{{{b}^{2}}-4c}) / 2$ of $\bm{G}$, with $b(\bm{q}) = - \mathrm{tr}(\bm{G})$ and $c(\bm{q}) = \mathrm{det}(\bm{G})$. The stability requirement for all wavevectors  $\operatorname{Re}[ \omega (\bm{q}) ] < 0$ is equivalent to, $b(\bm{q}) > 0$ and $c(\bm{q}) > 0$.
% In particular, our analysis shows that $\bm{G}$ is independent of $d$ and $\mu_d$. Thus, these two parameters will not result in instabilities or flows. 
% Further, since $b(\bm{q})$ and $c(\bm{q})$ are increasing functions of $\nu$ (Supplemental Material \cite{SM}, Sec. I), this self-propulsion parameter of dislocations tends to stabilize the smectic layers. 

We first consider the stability at small scales, i.e., $q \rightarrow +\infty$. At the leading order: $b( q\to \infty  ) \simeq ({{{\xi }_{m}}}/{{{m}_{H}}}){{q}^{4}}{{\cos }^{2}}\phi $ and $c( q\to \infty  ) \simeq ( {{{\xi }_{m}}}/{{{m}_{H}}}){{D}_{n}}{{q}^{6}}{{\cos }^{2}}\phi$ with $\phi = \arg(\bm{q})$ the argument of wavevector. This suggests that the active smectic system is always stable on small scales when $\xi_m > 0$. 
When $\xi_m = 0$, $b( q\to \infty ) \simeq [ {{D}_{n}}+{{D}_{m}} - (D_m + \widetilde{\lambda}_m / m_H) \cos^2\phi + (K/\eta) \cos^4\phi]{{q}^{2}}$ and $c( q\to \infty) \simeq {{D}_{n}} [ {{D}_{m}} - (D_m + \widetilde{\lambda}_m / m_H) \cos^2\phi + (K/\eta) \cos^4\phi]{{q}^{4}}$ with 
\begin{equation}
\widetilde{\lambda}_m = \lambda_m + 2 \nu_K \sqrt{\frac{\beta}{\alpha}} . 
\end{equation}
%$\widetilde{\lambda}_m = {{\lambda }_{m}}+{{\nu }_{K}}{{n}_{H}} = \lambda_m + 2 \nu_K \sqrt{\beta/\alpha}$. 
The stability requirement sets an upper limit of $\widetilde{\lambda}_m$ as $\widetilde{\lambda }_{m}^{\star} = {{m}_{H}}(2\sqrt{{{{D}_{m}}K}/{\eta }}-{{D}_{m}})$ if $D_m < K / \eta$ while $\widetilde{\lambda }_{m}^{\star}={K {{m}_{H}}}/{\eta }$ if $D_m > K / \eta$. 
When $\xi_m = 0$ and $\widetilde{\lambda}_m > \widetilde{\lambda}_m^{\star}$, the active smectic layers are unstable at small scales, leading to numerical instability.

\begin{figure}[t!]
\includegraphics[width=8.7cm]{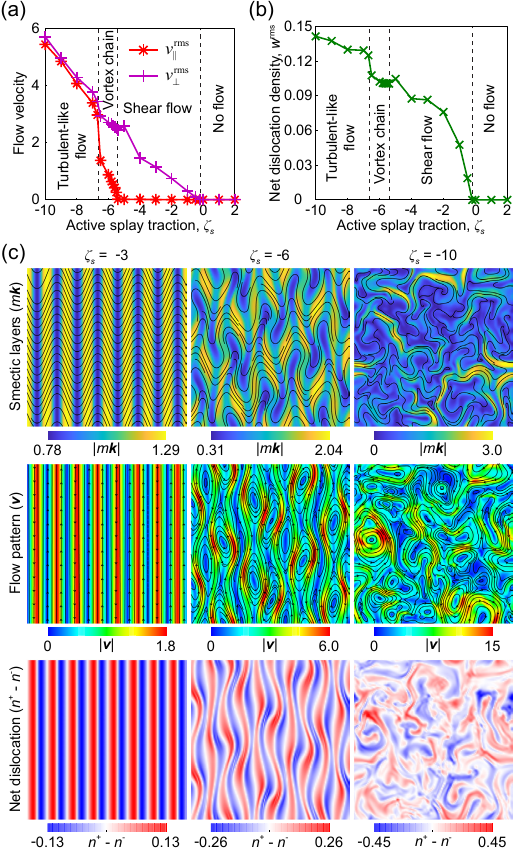}
\caption{Pattern formation driven by the active splay traction alone ($\nu_K = 0$).
(a, b) The flow velocities (a) and the net dislocation density (b) as functions of the active splay traction $\zeta_s$. Here, $v_{\parallel}^{\rm rms} = \sqrt{\langle v_x^2 \rangle}$, $v_{\perp}^{\rm rms} = \sqrt{\langle v_y^2 \rangle}$ and $w^{\rm rms} = \sqrt{\langle (n^{+} - n^{-})^2 \rangle}$. 
(c) Typical patterns at different levels of active splay tractions: (\textit{top}) smectic director, with black lines indicating the direction normal to the $\vec{k}$ field, and the color codes for the value of $m$; (\textit{middle}) flows, with arrows indicating the direction, and the color the velocity amplitude; (\textit{bottom}) net dislocation density. 
Other parameters set in Table 1. %, $\zeta_b = 0$, and $\beta = 0.1$.
}
\label{fig:Role_ActiveTraction}
\end{figure}

Next, we consider the stability at large scales, that is, $q \rightarrow 0$. In this limit, $b( q\to 0 ) \simeq \nu {{n}_{H}}>0$ and $c( q\to 0 ) \simeq \nu {{n}_{H}} ( {{D}_{n}}+{{g}_{1}}{{\cos }^{2}}\phi +{{g}_{2}}{{\cos }^{4}}\phi  ){{q}^{2}}$ with ${{g}_{1}} = -{{D}_{n}} -{{\widetilde{\lambda }_{m}}}/{{{m}_{H}}} - {{{\zeta }_{b}}}/{\lambda } - B( {{m}_{H}}-{{m}_{S}} )/{( \lambda {{m}_{H}} )}$ and ${{g}_{2}} = ( {{\zeta }_{s}}+{{\zeta }_{b}} )/{\lambda } + 2B( {{m}_{H}}-{{m}_{S}} )/{( \lambda {{m}_{H}} )}$. We find that the layers are stable along their normal direction ($\phi = {\pi}/{2}$) since $b>0$ and $c>0$ for $\phi = \pi / 2$. However, in general, the layers are not stable in all directions, as $c(q \rightarrow 0)<0$ for sufficiently large activity parameters ($\nu_K$, $\zeta_s$, $\zeta_b$, $\lvert m_S - m_H \lvert$). 

In particular, along the smectic layers ($\phi = 0$), the stability requirement results in 
\begin{equation}
\zeta _{s} > \zeta_s^{\rm cr}, \quad  \mathrm{with} \quad  \zeta_s^{\rm cr} = \min\{ \zeta_s^{\ast}, \zeta_s^{\diamond} \}, \label{eq:StabilityCondition_phi0}
\end{equation}
where $\zeta_s^{\ast} = B (m_S/m_H - 1) + {\lambda \widetilde{\lambda}_m} / {{{m}_{H}}}$ and ${\zeta}_s^{\diamond} = B (m_S/m_H - 1) + K\lambda/\eta + {{\xi }_{m}}{{\lambda }^{2}}/(\eta {{m}_{H}})$. 
Equation \eqref{eq:StabilityCondition_phi0} can reformulated into the following condition on the dislocation creation rate
\begin{equation}
\beta < \beta^{\text{cr}}, \quad  \mathrm{with} \quad \beta^{\text{cr}} =\frac{\alpha(\widetilde{\lambda}_m^{\rm cr} - \lambda_m)^2}{4{{\nu}_{K}^2}} , \label{eq:beta_critical}
\end{equation}
where $\widetilde{\lambda}_m^{\rm cr}$ is a critical value of $\widetilde{\lambda}_m$:
when ${{\zeta }_{s}} < {\zeta}_s^{\diamond}$, $\widetilde{\lambda}_m^{\rm cr} = \zeta_s m_H/\lambda + B (m_H - m_S)/\lambda$; otherwise, $\widetilde{\lambda}_m^{\rm cr} = (K m_H - \xi_m \lambda)/\eta + 2\sqrt{ \xi_m m_H (\zeta_s - {\zeta}_s^{\diamond})/\eta + \xi_m^2\lambda^2/\eta^2 }$ (see Supplemental Material \cite{SM}, Sec. I). 
%Thus, fast active creation of dislocations ($\beta > \beta^{\rm cr}$) can destabilize smectic layers. 
Equation \eqref{eq:beta_critical} indicates that the flow instability for $\beta > \beta^{\rm cr}$ is caused by the splay-induced motility of the dislocations ($\nu_K > 0$). In particular, when $\nu_K \rightarrow 0$, $\beta^{\rm cr} \rightarrow +\infty$, and the creation of dislocations cannot lead to steady-state flows. 
Equivalently, Eq. \eqref{eq:beta_critical} corresponds to a critical dislocation motility: $\nu_K^{\rm cr} = (\widetilde{\lambda}_m^{\rm cr} - \lambda_m) / n_H$. 
The case of $m_S \neq m_H$ can also lead to buckling instability when $m_S > m_{S}^{\rm cr}$, with 
\begin{equation}
m_{S}^{\text{cr}}={{m}_{H}}+\frac{{{\zeta }_{s}}{{m}_{H}}}{B}-\frac{\lambda }{B}\min \left\{ {{{\widetilde{\lambda }}}_{m}},\frac{K{{m}_{H}}+{{\xi }_{m}}\lambda }{\eta } \right\} , \label{eq:Critical_mS}
\end{equation}
obtained from Eq. \eqref{eq:StabilityCondition_phi0} (\cite{SM}). 
We check that $m_S^{\rm cr} = m_H$ for $\zeta_s = 0$ and $\widetilde{\lambda}_m = 0$.

\begin{figure}[t!]
\includegraphics[width=8.7cm]{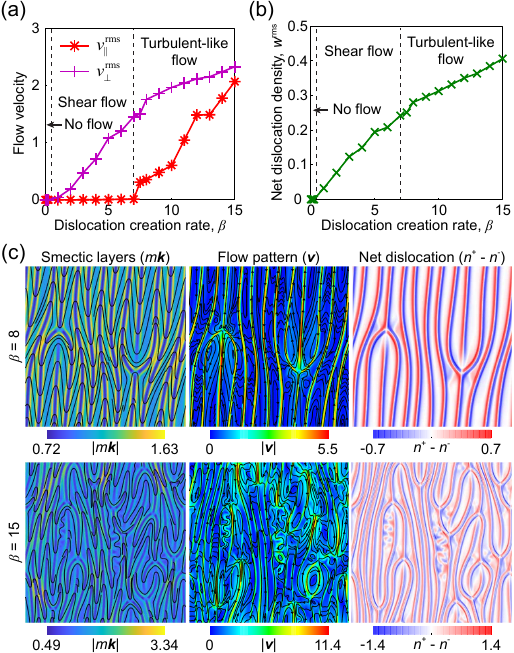}
\caption{
Pattern formation driven by the spontaneous creation of dislocations in the presence of splay-induced dislocation motility ($\nu_K = 1$), without active tractions. 
(a, b) The flow velocities (a) and the net dislocation density (b) as functions of the dislocation creation rate. 
% \jfbar{in the absence (symbols connected by dashed lines at $v=0$) and presence (symbols connected by solid lines) of a splay-induced activity at $\nu_K = \nu = 1$.}{}. 
% Here, $v_{\parallel}^{\rm rms} = \sqrt{\langle v_x^2 \rangle}$, $v_{\perp}^{\rm rms} = \sqrt{\langle v_y^2 \rangle}$ and $w^{\rm rms} = \sqrt{\langle (n^{+} - n^{-})^2 \rangle}$. 
(c) Typical patterns at different levels of dislocation creation rates: (\textit{left}) smectic layers; (\textit{middle}) flows; (\textit{right}) net dislocation density. Other parameters set in Table 1.}
\label{fig:beta_role}
\end{figure}
% Parameters: $\nu_K = 1$, and $\zeta_s = \zeta_b = 0$.

We then estimate the typical wavenumber $q_c$ for the buckling instability by maximizing $\mathrm{Re}[\omega_{+}(q)]$, with 
\begin{align*}
{{\omega }_{+}}=-\frac{{{\xi }_{m}}}{{{m}_{H}}}{{q}^{4}}-\frac{{{q}^{2}}[ m_H ({{\zeta }_{s}}-\zeta _{s}^{\text{cr}}) + ( K m_H-{{{{\widetilde{\lambda }}}_{m}}\eta} ){{q}^{2}} ]}{m_H( \lambda +\eta {{q}^{2}})} . \label{eq:omega_full}
\end{align*}
Near the transition, we obtain: 
\begin{equation}
q_c \approx \sqrt{\frac{{{m}_{H}}( \zeta _{s}^{\ast}-{{\zeta }_{s}})}{2( {{\xi }_{m}}\lambda +K{{m}_{H}}-\eta {{{\tilde{\lambda }}}_{m}})}} = \sqrt{\frac{\lambda ( \zeta _{s}^{\ast} - {{\zeta }_{s}} )}{2\eta ( {{\zeta }_{s}^{\diamond}} - \zeta _{s}^{\ast })}} , \label{eq:qc}
\end{equation}
with $\zeta _{s}^{\ast}$ and ${{\zeta }_{s}^{\diamond}}$  defined next to Eq. (\ref{eq:StabilityCondition_phi0}). An instability occurs as soon as $q_c$ is real, since $\omega_+(q_c)>\omega_+(0)=0$; we call it active Helfrich-Hurault instability, as in Ref. \cite{Kole_PRL_2021}, although the passive Helfrich--Hurault buckling occurs, in an infinite sample, at $q_c = 0$ \cite{Helfrich1971,Clark_APL_1973,Hurault1973}. In the vanishing friction limit ($\lambda \rightarrow 0$), 
%${{\omega }_{+}}=(\zeta_{s}^{\ast}-{{\zeta }_{s}}) / \eta -(K/\eta-\widetilde{\lambda}_m/m_H){{q}^{2}} -(\xi_m / m_H){{q}^{4}}$, which leads to 
% the most unstable wavevector reads 
${{q}^2_{c}}= \max \{\eta \widetilde{\lambda}_m-K {{m}_{H}}/({2{{\xi}_{m}}\eta}), 0 \}$; it is independent of the active traction $\zeta_s$, but the instability emerges for $\zeta_s < \zeta^{\mathrm{cr}}_s = B (m_S/m_H - 1)$. 

\begin{figure}[t!]
\includegraphics[width=8.7cm]{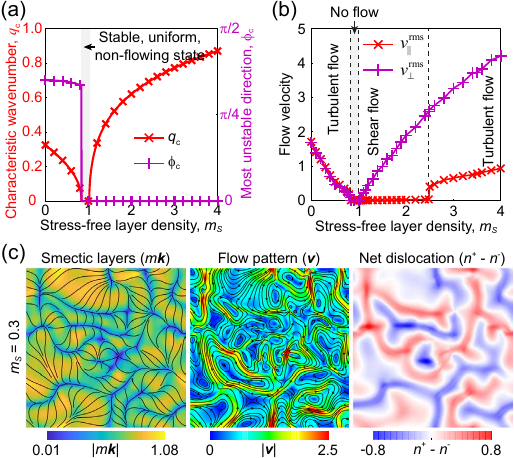}
\caption{Pattern formation driven by the stress-free layer spacing density $m_S$ alone. 
(a) The characteristic wavenumber $q_c$ and the most unstable direction $\phi_c$ as functions of the stress-free layer spacing density $m_S$. 
Here, $q_c$ and $\phi_c$ are determined by numerically exploring the maximum eigenvalue $\omega$ in the Fourier space $\bm{q}$. 
(b) The flow velocities as functions of the stress-free layer spacing density $m_S$. 
% Here, $v_{\parallel}^{\rm rms} = \sqrt{\langle v_x^2 \rangle}$ and $v_{\perp}^{\rm rms} = \sqrt{\langle v_y^2 \rangle}$. 
(c) Typical patterns at $m_S = 0.3$: (\textit{left}) smectic layers; (\textit{middle}) flows; (\textit{right}) net dislocation density. Parameters in Table 1 except for $\nu_K = 0$.}
\label{fig:Role_mS}
\end{figure}

\begin{table}[b]%The best place to locate the table environment is directly after its first reference in text
\caption{\label{tab:table1}%
Default set of parameters. 
}
\begin{ruledtabular}
\begin{tabular}{cccccccccccccccc}
$\lambda$   &
$\eta$      &
$D_m$       & 
$\lambda_m$   & 
$D_n$       & 
$m_S$       & 
$m_H$       & 
$B$         & 
$K$         & 
$\xi_m$     & 
$\alpha$    & 
$\beta$     & 
$\zeta_s$   &
$\zeta_b$   &
$\nu$       &
$\nu_K$     
\\
\colrule
0.2 & 1 & 1 & -1 & 1 & 1 & 1 & 10 & 5 & 0.5 & 1 & 0.1 & 0 & 0 & 1 & 1\\
\end{tabular}
\end{ruledtabular}
\end{table}

\textit{Numerical simulations.} -- 
We performed numerical simulations, verified the analytically determined threshold to flows, and explored the system behavior far beyond %the
threshold. %to flows.
%, (Fig. S1). 
We use a spectral method on a 256×256 two-dimensional lattice with periodic boundary conditions, a spacing $\Delta x = 0.25$, and a time step $\Delta t = 10^{-4}$, with weak perturbations to the homogeneous state $m\bm{k} = m_H \hat{\bm{y}}$ as the initial condition. 
If not stated otherwise, we consider the parameter values set in Table 1. %: $\lambda = 0.2$, $\eta = 1$, $D_m = 1$, $\lambda_m = -1$, $D_n = 1$, $m_S = m_H = 1$, $B = 10$, $K = 5$, $\xi_m = 0.5$, $\alpha = 1$, $\beta = 0.1$, $\zeta_s = \zeta_b = 0$, $\nu = 1$, and $\nu_K = 1$. 
%, $\chi = 0$, $\mu_d = 0$, and $d = 0$. 

\begin{figure}[t!]
\includegraphics[width=8.7cm]{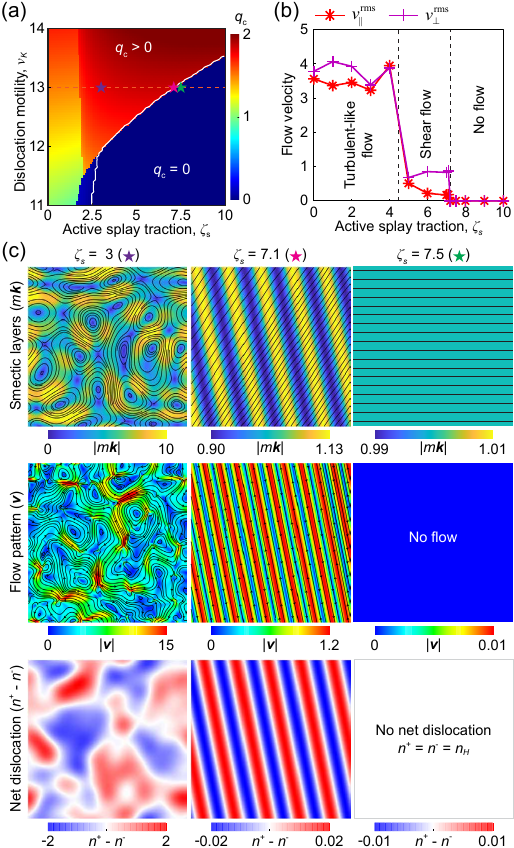}
\caption{Splay-induced motility of dislocations $\nu_K$ can lead to discontinuous transition to flows. 
(a) Analytical calculation of the characteristic wavenumber $q_c$ with respect to the active splay traction $\zeta_s$ and the splay-induced motility of dislocations $\nu_K$. 
The symbols correspond to simulations, as shown in (c). 
(b, c) Simulation results. 
(b) The flow velocity as a function of the active splay traction $\zeta_s$, where $\nu_K = 13$. 
(c) Typical patterns of active smectic layers at different levels of the active splay traction $\zeta_s$, where $\nu_K = 13$: (\textit{top}) smectic layers; (\textit{middle}) flows; (\textit{bottom}) net dislocation density. 
Other parameters set in Table 1.}
\label{fig:Simulation_DiscontinuousTransition}
\end{figure}

Decreasing $\zeta_s$ below the critical value $\zeta_s^{\rm cr} = -0.2$ (Eq. (\ref{eq:StabilityCondition_phi0})), results in buckling and shear flows of the smectic layers (Fig. \ref{fig:Role_ActiveTraction}(a,c)), at the wavenumber predicted by Eq. (\ref{eq:qc}) (Fig. S1). Further decreasing $\zeta_s$ below a second threshold $\zeta_s^{\rm cr,2} \approx -5.4$ results in a stable vortex chain pattern (Fig. \ref{fig:Role_ActiveTraction}(a,c)). Beyond a third threshold $\zeta_{s}^{\rm cr,3} \approx -6.7$, the system transits to turbulent-like flows, with an increasing number of vortices, as expected from Eq. \eqref{eq:qc}, see Fig. \ref{fig:Role_ActiveTraction}(a, c). 

For nonzero splay-induced dislocation motility ($\nu_K > 0$), we find that the increase of the dislocation creation rate $\beta$ (or of the splay-induced dislocation motility $\nu_K$) beyond $\beta^{\rm cr}$ (or $\nu_K^{\rm cr}$) can also lead to stable shear flows; while higher $\beta$ (or $\nu_K$) results in turbulent-like flows (Fig. \ref{fig:beta_role} and SM, Fig. S3 \cite{SM}). 
% When $\nu = 0$, a large value of $\beta$ can even lead to a new equilibrium homogeneous state ($m_{\rm eq} > m_H$) with more layers (Fig. S6). 

The compression-induced dislocation motility $\nu$ does not lead to flows when all other active parameters are set to zero (e.g., $\zeta_b = 0$) but does when $\zeta_b > 0$, eventually leading to turbulent-like flows (see SM, Fig. S4 \cite{SM}). 

A homeostatic interlayer spacing different from the rest ($m_S \neq m_H$) can also drive flows; see Fig. \ref{fig:Role_mS}. Increasing $m_S > m_S^{\rm cr}$ (Eq. \eqref{eq:Critical_mS}) leads to stable shear flows and eventually turbulent-like flows. Decreasing $m_S < m_H$, the most unstable direction is tilted with respect to smectic layers (Fig. S2 \cite{SM}); a uniform, non-flowing state transitions directly to turbulent-like flows, see Fig. \ref{fig:Role_mS}(c). 

Combining the active traction and splay-induced motility, we find that the most unstable wavevector $q_c$ undergoes a discontinuous transition (first-order-like) from zero to a finite value when $\zeta_s^{\diamond} < \zeta_s^{\ast}$, i.e., $\widetilde{\lambda}_m > (K m_H + \xi_m \lambda) / \eta$ (Fig. \ref{fig:Simulation_DiscontinuousTransition}(a) and Fig. S5 \cite{SM}). Such behavior is validated by simulations: the quiescent state ($v^{\mathrm{rms}}=0$) transitions into a turbulent-like regime with a finite velocity $v^{\mathrm{rms}}>0$ at the predicted critical activity (Fig. \ref{fig:Simulation_DiscontinuousTransition}). Such a velocity jump upon varying activity was recently observed in active nematics \cite{Bell_PRL_2022}, contrasting with the continuous transitions predicted so far \cite{Voituriez2005}.

\textit{Dislocation free case.} -- 
We consider the case of no dislocations $n^{\pm} = 0$ in the Supplemental Material \cite{SM}, Sec. II. We also observe stable shear flow, localized vortex chain, and turbulent-like flows (Fig. S7 \cite{SM}). 
Compared to the case with active dislocations, the transition from a non-flowing homogeneous state to a shear flow pattern occurs at a higher active traction threshold, with $\Delta \zeta_s^{\rm cr} = - \lambda \nu_K n_H / m_H$, and we did not find signs of any discontinuous transition to flows. %(Supplemental Material \cite{SM}, Sec. II). 

\textit{Perspectives} -- As we focused on the leading order terms in the gradient expansion, we have ignored the Ericksen stress, which itself can lead to flows (see the magical spiral, page 158 in Ref. \cite{de1993physics}). A proper thermal equilibrium could be reached by including the Ericksen stress while imposing $m_S = m_H$, $\lambda_m = 0$, and with $\xi_m$ and $D_m$ fixed by $B$ and $K$ (Supplemental Material \cite{SM}, Sec. II). 

\textit{Conclusion.} -- In this Letter, we established an active hydrodynamic theory for active smectic layers that includes dislocations in two-dimensional space. We look forward to applying our results to flows in actomyosin layers within fibroblast cells \cite{Hu2017}.
%, or the structure of the endoplasmic reticulum. 

\textit{Acknowledgements} -- 
The project leading to this publication has received funding from France 2030, the French Government program managed by the French National Research Agency (ANR-16-CONV-0001) and from the Excellence Initiative of Aix-Marseille University - A*MIDEX, ANR-20-CE30-0023 (COVFEFE) and ANR-21-CE13-0050 (CODAC). 

%\bibliographystyle{apsrev4-1}
%\bibliography{refs}


\end{document}